\documentclass[nofootinbib,preprint,preprintnumbers,a4paper,11pt]{revtex4}
\usepackage{epsfig,footnote}
\usepackage{ulem}
\usepackage{color}
\usepackage{array}
\usepackage{amssymb}
\usepackage{amsmath}
\usepackage{graphicx,subfigure}
\usepackage{longtable}
\usepackage{verbatim}
\usepackage{amsfonts}
\usepackage{hyperref}
\usepackage{multirow}

\begin{document}

\title{$K_S^0-K_L^0$ Asymmetries and $CP$ Violation\\ in Charmed Baryon Decays into  Neutral Kaons}

\author{Di Wang, Peng-Fei Guo, Wen-Hui Long and
Fu-Sheng Yu\footnote{Corresponding author.}\footnote{Email: yufsh@lzu.edu.cn} }

\address{School of Nuclear Science and Technology,  Lanzhou University,  Lanzhou 730000,  China}

\begin{abstract}

We study the $K^0_S-K^0_L$ asymmetries and $CP$ violations in charm-baryon decays with neutral kaons in the final state.
The $K^0_S-K^0_L$ asymmetry can be used to search for two-body doubly Cabibbo-suppressed amplitudes of charm-baryon decays, with the one in $\Lambda^+_c\to pK^0_{S,L}$ as a promising observable.
Besides, it is studied for a new $CP$-violation effect in these processes, induced by the interference between the Cabibbo-favored and doubly Cabibbo-suppressed amplitudes with the neutral kaon mixing.
Once the new CP-violation effect is determined by experiments, the direct $CP$ asymmetry in neutral kaon modes can then be extracted and used to search for new physics. The numerical results based on $SU(3)$ symmetry will be tested by the experiments in the future.
\end{abstract}

\maketitle

\section{Introduction}

Charm physics plays an important role in studying the perturbative and non-perturbative QCD and searching for new physics with special structure in the up-type quark sector. Unlike the charmed meson decays with fruitful results during the past decades \cite{Artuso:2008vf}, the study of weak decays of charmed baryons has been made little progress  both in theory and in experiment until a few years ago when some new measurements were performed by the Belle and BESIII experiments \cite{Pal:2017ypp,Ablikim:2016vqd,Ablikim:2017iqd,Ablikim:2016tze,Ablikim:2017ors,Zupanc:2013iki,Ablikim:2015prg,Ablikim:2015flg,Yang:2015ytm,Ablikim:2016mcr}.
Charmed baryon physics is becoming intriguing with more data available and collected by Belle (II), BESIII and LHCb. Charmed baryon decays provide an ideal laboratory to study the heavy-to-light baryonic transitions \cite{Korner:1992wi,Korner:1978tc,Ivanov:1997ra,Cheng:2015rra,Cheng:1991sn,Zenczykowski:1993hw,Uppal:1994pt,Fayyazuddin:1996iy} and test the flavor $SU(3)$ symmetry \cite{Lu:2016ogy,Savage:1989qr,Kohara:1991ug,Verma:1995dk,Chau:1995gk,Sharma:1996sc,Savage:1991wu,Geng:2017esc}.
They also provide the essential inputs for the decays of $b$-flavored hadrons decaying into charmed baryons to determine the CKM matrix element $|V_{cb}|$ \cite{Detmold:2015aaa,Dutta:2015ueb}.
In this work, we will study charmed baryon decays into neutral kaons.

Under the flavor $SU(3)$ symmetry, the ground states of $\Lambda^+_c$, $\Xi^+_c$ and $\Xi^0_c$ form an anti-triplet, all of which decay weakly.
In the Standard Model (SM), the charmed baryon weak decays are classified into three types: the Cabibbo-favored (CF) decays, the singly Cabibbo-suppressed (SCS) decays and the doubly Cabibbo suppressed (DCS) decays \cite{Patrignani:2016xqp}.
It can be generally expected that the branching fractions of the CF, SCS and DCS modes are of the order of $10^{-2}$, $10^{-3}$ and $10^{-4}$, respectively.
Due to the small branching fractions, the DCS decays are more difficult to be observed compared to the CF and SCS decays.
The first and only evidence of DCS transitions of charmed baryons is found in a three-body $\Lambda_c^+$ decay by the Belle collaboration \cite{Yang:2015ytm},
\begin{equation}
  BR(\Lambda_c^+\to pK^+\pi^-)/ BR(\Lambda_c^+\to pK^-\pi^+)=(2.35\pm0.27\pm0.21)\times 10^{-3}.
\end{equation}
But none of the two-body DCS processes has been observed to date. The two-body DCS decays play an essential role in understanding the dynamics of charmed hadron decays, since the multi-body decays are difficult to be studied in theory. Besides, new physics might contribute to the relatively small DCS amplitude, leading to a much larger direct $CP$ violation compared to the SM prediction \cite{Lipkin:1999qz,Xing:1995jg,Bigi:1994aw,DAmbrosio:2001mpr}. Therefore, it is important to search for the two-body DCS amplitudes of charm-baryon decays. Except for the direct measurement on the DCS processes, the two-body DCS amplitudes can also be revealed by the $K^0_S-K^0_L$ asymmetry in charmed baryon decays into neutral kaons. The $K^0_S-K^0_L$ asymmetry is induced by the interference between the CF and DCS amplitudes, which has been studied in $D$ meson decays \cite{Muller:2015lua,Gao:2014ena,Gao:2006nb,Bhattacharya:2009ps,Cheng:2010ry,Bigi:1994aw,Wang:2017ksn,He:2007aj}. In this work, we investigate the $K^0_S-K^0_L$ asymmetry in charmed baryon decays, and find that  the one in $\Lambda_c^+\to pK_{S,L}^0$ is the promising observable to search for two-body DCS charm-baryon decay amplitudes.

$CP$ violation can occur in charmed baryon decays into neutral kaons. It is an essential element to interpret the matter-antimatter asymmetry in the Universe \cite{Sakharov:1967dj} and provides a window to search for new physics beyond the SM.
$CP$ asymmetries have been well established in kaon and $B$ meson systems \cite{Patrignani:2016xqp}.
In the baryon sector, the only signal of $CP$ asymmetry is found in $\Lambda_b^0\to p\pi^-\pi^+\pi^+$ with  $3.3\,\sigma$ \cite{Aaij:2016cla}. $CP$ violation in charmed baryon decays has not been observed up to now. 
In some literatures, the $CP$ violation in charm decays into neutral kaons have be studied \cite{Azimov:1998sz,Amorim:1998pi,DAmbrosio:2001mpr,Ko:2012pe,Lipkin:1999qz,Grossman:2011zk,Bianco:2003vb}.
In Ref. \cite{Yu:2017oky}, a new measurable $CP$-violation effect is found existing
in $D$ meson decays into neutral kaons,
except for the known indirect $CP$ violation in $K^{0}-\overline K^{0}$ mixing and direct $CP$ asymmetry in charm decays. The new effect is induced by the interference between the CF and DCS amplitudes with the mixing of final-state mesons.
In this work, we will show that the new $CP$-violation effect also exist in charmed baryon decays.
Once the new effect was well determined in experiment, the direct $CP$ asymmetry could be obtained and used to search for new physics.

Numerically, the dynamics of charmed baryon decays is always difficult to describe, due to the sizable non-factorizable contributions \cite{Korner:1992wi,Korner:1978tc,Ivanov:1997ra,Cheng:2015rra,Cheng:1991sn,Zenczykowski:1993hw,Uppal:1994pt,Fayyazuddin:1996iy,Lu:2016ogy,Savage:1989qr,Kohara:1991ug,Verma:1995dk,Chau:1995gk,Sharma:1996sc,Savage:1991wu,Geng:2017esc}. In order to estimate the $K^0_S-K^0_L$ asymmetry and $CP$ asymmetry, we analyze the decays of charm-baryon anti-triplet into light baryon octet and pseudoscalar octet based on the flavor $SU(3)$ symmetry, in which
those universal parameters are extracted from the available data.
Our results are well consistent with the measured data and can be tested by Belle II, BESIII and LHCb.

This paper is organized as follows. In Sec. \ref{th1}, we discuss the $K^0_S-K^0_L$ asymmetry in charmed baryon decays into neutral kaons and its search for the DCS transitions. The corresponding time-dependent and time-integrated $CP$ asymmetries are studied in Sec. \ref{th2}.
The numerical analysis is given in Sec. \ref{re}.
And Sec. \ref{co} is the summary.

\section{$K^0_S-K^0_L$ asymmetry}\label{th1}

In the charmed hadron decays into neutral kaons, the interference between the Cabibbo-favored (CF) and the doubly Cabibbo-suppressed (DCS) amplitudes leads to the $K^0_S-K^0_L$ asymmetry, an observable to search for the DCS transitions.
Specifically, the $K_S^0-K_L^0$ asymmetry in charmed baryon $(\mathcal{B}_c)$ decaying into light baryon $(\mathcal{B})$ and neutral kaons is defined as
 \begin{equation}\label{a1}
   R(\mathcal{B}_c\to \mathcal{B}K^0_{S,L})\equiv\frac{\Gamma(\mathcal{B}_c\rightarrow \mathcal{B}K_S^0) -\Gamma(\mathcal{B}_c\rightarrow \mathcal{ B}K_L^0)}{\Gamma(\mathcal{B}_c\rightarrow \mathcal{B}K_S^0) + \Gamma(\mathcal{B}_c\rightarrow \mathcal{B}K_L^0)}.
 \end{equation}
Due to the fact that the $K^0_S-K^0_L$ asymmetry is not sensitive to the $CP$ violating effect in the $K^0-\overline K^0$ mixing \cite{Wang:2017ksn}, the $K^0_S$ and $K^0_L$ states can be referred as the $CP$ eigenstates
\begin{equation}~\label{eq:KSKL11}
|K_{S}^0\rangle  =   \frac{1}{\sqrt{2}}\left(|K^0\rangle-|\overline{K}^0\rangle\right), \qquad
|K_{L}^0\rangle  =   \frac{1}{\sqrt{2}}\left(|K^0\rangle+|\overline{K}^0\rangle\right),
 \end{equation}
under the convention $\mathcal{CP}|K^0\rangle=-|\overline K^0\rangle$.

The CF amplitude of $\mathcal{B}_c\to \mathcal{B}\overline K^0$ and the DCS amplitude of $\mathcal{B}_c\to  \mathcal{B}K^0$ decays read as
\begin{equation}
\mathcal{A}(\mathcal{B}_c\to \mathcal{B}\overline K^0)=\mathcal{T}_{\rm CF}\, e^{i(\phi_{\rm CF}+\delta_{\rm CF})},\qquad
\mathcal{A}(\mathcal{B}_c\to  \mathcal{B}K^0)=\mathcal{T}_{\rm DCS}\, e^{i(\phi_{\rm DCS}+\delta_{\rm DCS})},
\end{equation}
where $\mathcal{T}_{\rm CF}~(\mathcal{T}_{\rm DCS})$ is the magnitude of the CF (DCS) amplitude, and $\delta_{\rm CF}~(\delta_{\rm DCS})$ and $\phi_{\rm CF}~(\phi_{\rm DCS})$ are the relative strong and weak phases, respectively.
The decay amplitudes of $\mathcal{B}_c\to  \mathcal{B}K^0_S$ and $\mathcal{B}_c\to  \mathcal{B}K^0_L$ are then
\begin{equation}\label{eq:ampKSKL}
\begin{split}
 \mathcal{A}(\mathcal{B}_c\to  \mathcal{B}K^0_S)  &   =
\frac{1}{\sqrt{2}}\mathcal{T}_{\rm DCS}\, e^{i(\phi_{\rm DCS}+\delta_{\rm DCS})}
- \frac{1}{\sqrt{2}}\mathcal{T}_{\rm CF}\, e^{i(\phi_{\rm CF}+\delta_{\rm CF})}, \\
\mathcal{A}(\mathcal{B}_c\to  \mathcal{B}K^0_L)  &   =  \frac{1}{\sqrt{2}}\mathcal{T}_{\rm DCS}\, e^{i(\phi_{\rm DCS}+\delta_{\rm DCS})} +\frac{1}{\sqrt{2}}\mathcal{T}_{\rm CF}\, e^{i(\phi_{\rm CF}+\delta_{\rm CF})}.
\end{split}
\end{equation}
Similar as Ref. \cite{Wang:2017ksn}, the ratio between the DCS and CF amplitudes are defined as
\begin{equation}\label{a3}
 {\mathcal{A}(\mathcal{B}_c\to \mathcal{B}K^0)\over \mathcal{A}(\mathcal{B}_c\to\mathcal{B}\overline K^0)} \equiv
  r_f e^{i(\phi+\delta_f)},
\end{equation}
where $r_f=\mathcal{T}_{\rm DCS}/\mathcal{T}_{\rm CF}$, $\phi=\phi_{\rm DCS} - \phi_{\rm CF}$ and $\delta_f=\delta_{\rm DCS} - \delta_{\rm CF}$.
The parameters $r_f$ and $\delta_f$ depend on the individual processes, while $\phi$ is mode independent in the SM.
It is found that if the DCS amplitude vanishes, $r_f=0$.
In the SM, $r_f$ is expected to be proportional to the ratio $|V_{cd}^*V_{us}/V_{cs}^*V_{ud}| \sim \lambda^2 \sim \mathcal{O}(10^{-2})$, and the weak phase is negligibly small, i.e., $\phi\equiv Arg[-V_{cd}^*V_{us}/V_{cs}^*V_{ud}]=(-6.2\pm 0.4)\times 10^{-4}$.
The $K^0_S-K^0_L$ asymmetry can be further reduced as \cite{Wang:2017ksn}
\begin{equation}\label{x4}
   R(\mathcal{B}_c\to \mathcal{B}K^0_{S,L})={|\,1-\,r_f\,e^{i\delta_f}|^{2}-|1\,+\,r_f\,e^{i\delta_f}|^{2}\over
|\,1-\,r_f\,e^{\delta_f}|^{2}+|1\,+\,r_f\,e^{i\delta_f}|^{2}}  \simeq -2r_f\cos\delta_f,
\end{equation}
which is expected to be of the order of $\mathcal{O}(10^{-2})$, being measurable in experiments.

The $K^0_S-K^0_L$ asymmetry could be used to search for the DCS decays in charmed baryon decays. If the DCS transition is absent, i.e., $r_{f}=0$, the $K^0_S-K^0_L$ asymmetry shall vanish.
Thereby,
a non-zero experimental result of $R(\mathcal{B}_c\to \mathcal{B}K^0_{S,L})$, namely $r_{f}\neq0$, can be taken as the evidence for the DCS transitions of charmed baryons.
In fact, since the branching fraction of $\Lambda_c^+\to p K^0_S$ is measured as \cite{Ablikim:2015flg}
\begin{equation}
  \mathcal{B}r(\Lambda_c^+\to pK^0_S)_{\text{exp}}=(1.52\pm0.08\pm0.03)\%,
\end{equation}
it can be expected that $\Lambda_c^+\to pK^0_L$ has a branching fraction of the order of one percent, and thus might be measured with large data sample. Therefore, the corresponding $K^0_S-K^0_L$ asymmetry can reveal the DCS amplitude of $\Lambda^+_c\to pK^0$.

In the following discussions, we will see that the $K_S^0-K_L^0$ asymmetry in the decays of $\Lambda_c^+\to p K_{S,L}^0$ is a promising observable to search for the two-body DCS amplitude of charmed baryon decays.
At BESIII, only $\Lambda_c^+$ can be produced due to the energy limit. At Belle (II) and LHCb, the productions of $\Xi^+_c$ and $\Xi^0_c$ with an additional strange quark are smaller than $\Lambda^+_c$.
For the two-body $\Lambda^+_c$ decays into the light baryon octet and a pseudoscalar meson, the only two DCS modes are $\Lambda^+_c\to pK^0$ and $\Lambda^+_c\to nK^+$.
In the latter mode, neutron is always difficult to be detected in experiments. On the contrary, $\Lambda^+_c\to pK^0$ can be revealed by the $K_S^0-K_L^0$ asymmetry in $\Lambda_c^+\to pK_{S,L}^0$ with $K_L^0$ detected with a high efficiency at Belle (II).
In the decay modes of $\Lambda^+_c$ into the baryon octet and a vector meson, and the baryon decuplet and a pseudoscalar meson, the DCS transitions include $\Lambda^+_c\to pK^{*0}$, $nK^{*+}$, $\Delta^+K^0$, $\Delta^0K^+$.
Among them, the decays of $\Lambda^+_c\to nK^{*+}(\to K^+\pi^0)$ and $\Lambda^+_c\to K^0\Delta^+(\to p\pi^0)$ suffer from the low efficiencies of detections of neutrons or $\pi^0$.
The modes of $\Lambda^+_c\to pK^{*0}(\to K^+\pi^-)$ and $\Lambda^+_c\to K^+\Delta^0(\to p\pi^-)$ are actually included in the measured three-body DCS decay $\Lambda^+_c\to pK^+\pi^-$ \cite{Yang:2015ytm}, while a partial-wave analysis requires much more data.
In short, the $K_S^0-K_L^0$ asymmetry in $\Lambda_c^+\to pK_{S,L}^0$ decays is of a priority to investigate the two-body DCS charm-baryon-decay amplitudes in experiments, especially at Belle (II).

\section{$CP$ asymmetry}\label{th2}

$CP$ violation can occur in the charm decays into neutral kaons, induced by the interference between the CF and DCS amplitudes with the $K^{0}-\overline K^{0}$ mixing. As pointed out in \cite{Yu:2017oky}, there exist three $CP$-violation effects in charmed meson decays, i.e., the indirect $CP$ violation in $K^{0}-\overline K^{0}$ mixing, the direct $CP$ asymmetry in charm decays, and the effect from the interference between two tree (CF and DCS) amplitudes with neutral kaon mixing. It is also worthwhile to study the $CP$-violation effects in the charmed baryon decays.

Unlike Eq. (\ref{eq:KSKL11}),  the indirect $CP$ violation in $K^{0}-\overline K^{0}$ mixing should be taken into account for the study of $CP$ violation effects in charmed baryon decays. Such that
the $K_S^0$ and $K_L^0$ states are
\begin{equation}\label{eq:KSKL}
\begin{split}
|K_{S,L}^0\rangle & =   p|K^0\rangle\mp q|\overline{K}^0\rangle,
\end{split}
 \end{equation}
where
$p = (1+\epsilon)/{\sqrt{2(1+|\epsilon|^2)}}$,
$q =  (1-\epsilon)/{\sqrt{2(1+|\epsilon|^2)}},$
and $\epsilon$ is a small complex parameter characterizing the indirect $CP$ asymmetry in the $K^0-\overline K^0$ mixing system, with the values of $|\epsilon|=(2.228\pm0.011)\times 10^{-3}$ and its phase $\phi_\epsilon=43.52^\circ\pm0.05^\circ$ \cite{Patrignani:2016xqp}.
In experiments, the $K_{S}^{0}$ state is actually reconstructed by $\pi^{+}\pi^{-}$.  The time-dependent $CP$ asymmetry in the decay chain of $\mathcal{B}_c\to \mathcal{B}K(t)(\to \pi^+\pi^-)$ reads as
\begin{equation}\label{m1}
A_{CP}(t) \equiv\frac{\Gamma_{\pi\pi}(t)-\overline
\Gamma_{\pi\pi}(t)}{\Gamma_{\pi\pi}(t)+\overline\Gamma_{\pi\pi}(t)},
\end{equation}
with $  \Gamma_{\pi\pi}(t)\equiv\Gamma(\mathcal{B}_c\to \mathcal{B}K(t)(\to \pi^{+}\pi^{-}))$
and $\overline\Gamma_{\pi\pi}(t)\equiv\Gamma(\overline {\mathcal{B}}_c\to \overline {\mathcal{B}}K(t)
(\to \pi^{+}\pi^{-}))$,
where the intermediate state $K(t)$ is recognized as a time-evolved neutral kaon $K^0(t)$ or $\overline K^0(t)$, and $t$ is the time difference between the charmed baryon decays and the neutral kaon decays in the kaon rest frame  \cite{Grossman:2011zk,Yu:2017oky}.
 The amplitude of $ \mathcal{A}(\mathcal{B}_c\to \mathcal{B}K(t)(\to \pi^+\pi^-))$ can be deduced as
\begin{align}
 \mathcal{A}(\mathcal{B}_c\to \mathcal{B}K(t)(\to& \pi^+\pi^-))
   =  \mathcal{A}(\mathcal{B}_c\rightarrow \mathcal{B}K^0)\big[g_+(t)\mathcal{A}(K^0\to \pi^+\pi^-)+\frac{q}{p}g_-(t)\mathcal{A}(\overline K^0\to \pi^+\pi^-)\big]
   \nonumber\\
 & +\mathcal{A}(\mathcal{B}_c\rightarrow \mathcal{B}\overline K^0)\big[g_+(t)\mathcal{A}(\overline K^0\to \pi^+\pi^-)+\frac{p}{q}g_-(t)\mathcal{A}(K^0\to \pi^+\pi^-)\big],
\end{align}
in which $g_+$ and $g_-$ describe the flavor preserving and flavor changing time evolutions, respectively,
\begin{equation}
  g_\pm(t)=\frac{1}{2}e^{-i (m_L -\frac{i}{2}\Gamma_L)t}\pm \frac{1}{2}e^{-i (m_S -\frac{i}{2}\Gamma_S)t},
\end{equation}
with the mass $m_{S}$ ($m_L$) and the width $\Gamma_{S}$ ($\Gamma_L$) of the $K^0_S$ ($K^0_L$) meson.

\begin{figure}[t!]
\centering
\includegraphics[scale=0.30]{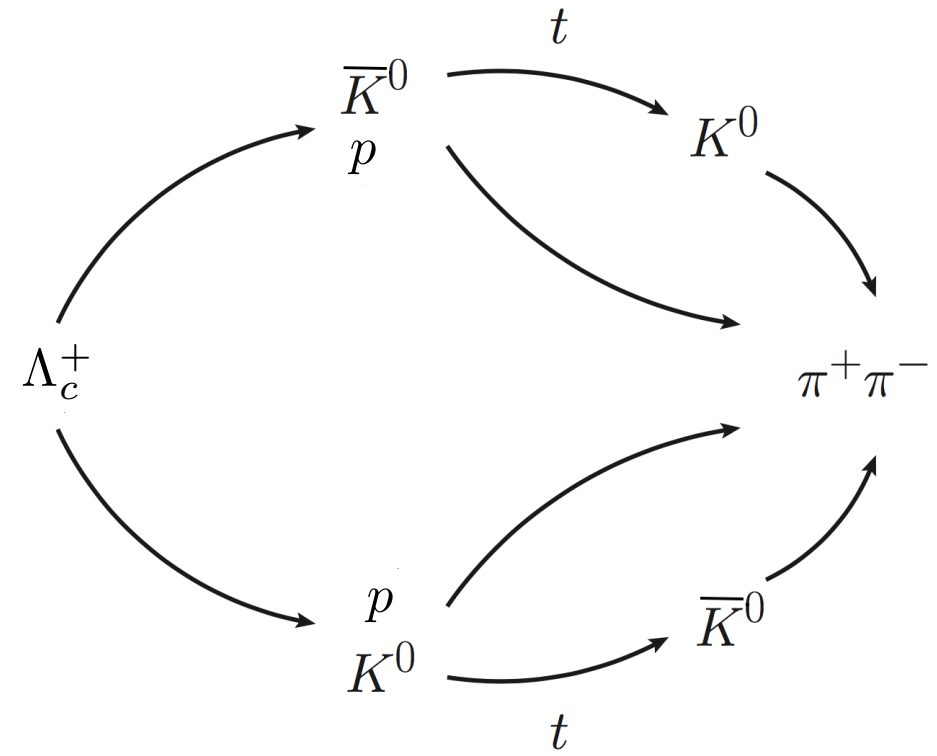}
\caption{Schematic description of the chain decay $\Lambda^+_c\to pK(t)(\to \pi^+\pi^-)$.
} \label{fig:amp}
\end{figure}

Neglecting the tiny direct $CP$ asymmetry in $K^0\to \pi^+\pi^-$, i.e., $\mathcal{A}(\overline K^0 \to \pi^+\pi^-)=-\mathcal{A}(K^0 \to \pi^+\pi^-)$, the time-dependent $CP$ asymmetry is approximated as
\begin{equation}\label{eq:KSAcp}
A_{CP}(t)\simeq\big[A_{CP}^{\overline K^0}(t)+A_{CP}^{\text{dir}}(t)+A_{CP}^{\text{int}}(t)\big]/D(t),
\end{equation}
with
$D(t)= e^{-\Gamma_St}(1-2\,r_f\cos\delta_f\cos\phi)+e^{-\Gamma_Lt}|\epsilon|^2$.
The first term $A_{CP}^{\overline K^0}(t)$ in the numerator denotes the $CP$ violation in neural kaon mixing \cite{Grossman:2011zk},
\begin{align}\label{q1}
A_{CP}^{\overline K^0}(t)
&=  2\mathcal{R}e[\epsilon]e^{-\Gamma_St}-2e^{-\Gamma t}
\big(\mathcal{R}e[\epsilon]\cos(\Delta mt)+\mathcal{I}m[\epsilon]\sin(\Delta mt)\big),
\end{align}
with $\Gamma\equiv(\Gamma_S+\Gamma_L)/2$, $\Delta m\equiv m_L-m_S$. It is clear that $A_{CP}^{\overline K^0}(t)$ is independent of $r_f$, i.e., independent on the DCS amplitude.
The second term $A_{CP}^{\text{dir}}(t)$ is the direct $CP$ asymmetry induced by the interference between the CF and DCS amplitudes,
\begin{align}\label{q2}
A_{CP}^{\text{dir}}(t)=2e^{-\Gamma_St}\,r_f\sin\delta_f\sin\phi.
 \end{align}
The third term $A_{CP}^{\text{int}}(t)$ originates from the interference between the two tree (CF and DCS) amplitudes with the neutral kaon mixing,
\begin{align}\label{q3}
A_{CP}^{\text{int}}(t)
&= -4\,r_f\cos\phi\sin\delta_f\big(\mathcal{I}m[\epsilon]e^{-\Gamma_St}-e^{-\Gamma t}
(\mathcal{I}m[\epsilon]\cos(\Delta mt)-\mathcal{R}e[\epsilon]\sin(\Delta mt))\big).
 \end{align}
 $A_{CP}^{\text{int}}(t)$ has been missed in the literature of studying $D$ meson decays \cite{DAmbrosio:2001mpr,Ko:2012pe,Lipkin:1999qz,Grossman:2011zk,Bianco:2003vb} in which the DCS amplitudes are assumed to be zero, and was first pointed out in \cite{Yu:2017oky} as a new $CP$-violation effect. It arises from the mother decay and the
daughter mixing. Its physical meaning can be depicted in Fig.~\ref{fig:amp}, a schematic description of the chain decay, taking $\Lambda^+_c\to pK(t)(\to \pi^+\pi^-)$ as an example.
$A_{CP}^{\rm int}(t)$ is from the interference effect between the Cabibbo-favored amplitude of $\Lambda_c^+\to p \overline K^0$ with neutral kaon mixing $\overline K^0\to K^0(t)\to \pi^+\pi^-$, and the doubly Cabibbo-suppressed amplitude of $\Lambda_c^+\to p K^0$ without kaon mixing $K^0\to K^0(t)\to \pi^+\pi^-$. Equivalently, the new $CP$-violation effect is also from the interference effect between the amplitudes of $\Lambda_c^+\to p \overline K^0$ without kaon mixing $\overline K^0\to \overline K^0(t)\to \pi^+\pi^-$, and $\Lambda_c^+\to p K^0$ with kaon mixing $K^0\to \overline K^0(t)\to \pi^+\pi^-$. The weak phase of $A_{CP}^{\rm int}(t)$ is from the (daughter) kaon mixing, $\epsilon$, while the strong phase is from the (mother) charm decays, $\delta_f$. Its mechanism is more complicated
than for the ordinary mixing-induced $CP$ asymmetry in,
for instance, the $D^0(t)\to K^+K^-$ mode, in which both the oscillation and decay take place in the mother particle $D^0(t)$. Besides, $A_{CP}^{\text{int}}(t)$ isn't the direct $CP$ asymmetry in charm decays, since it doesn't
vanish as $\phi\rightarrow 0$. In order to investigate the direct $CP$ violation $A_{CP}^{\rm dir}(t)$ which is tiny in the SM due to the smallness of $\phi$ and thus sensitive to new physics, it is necessary to extract $A_{CP}^{\overline K^0}(t)$ and $A_{CP}^{\rm int}(t)$ from the total $CP$ violation, seen in Eq. \eqref{eq:KSAcp}. Thus it is worthwhile to study $A_{CP}^{\rm int}(t)$ in the relevant processes.
In Eq.~(\ref{q3}), neglecting the weak phase in charm, i.e., setting $\phi\to 0$, only $r_f$ and $\delta_f$ are required to predict the values of $A_{CP}^{\rm int}(t)$, since $\epsilon$ is well determined in experiment. Fortunately, $r_f$ and $\delta_f$ can be obtained by the data of branching fractions, which will be discussed in Section IV.

From Eqs. \eqref{q1} $\sim$ \eqref{q3}, it is found that $A_{CP}^{\overline K^0}(t)$ and $A_{CP}^{\text{int}}(t)$ vanish at $t=0$ and
\begin{equation}
  A_{CP}(t=0) = A_{CP}^{\text{dir}}(t=0) = 2\,r_f\sin\delta_f\sin\phi.
\end{equation}
In the SM, $A_{CP}(t=0)$ is of order of $10^{-5}$ since $r_f\sim \mathcal{O}(10^{-2})$ and $\phi\sim \mathcal{O}(10^{-4})$.
Such an order is far beyond the precision limit of the forthcoming experiments, Belle II and LHCb upgrade.
However, in some new physics models, the weak phase difference can be large, which results in a larger $A_{CP}(t=0)$.
Thereby, an observation with nonvanishing $A_{CP}(t=0)$ would be the signal of new physics \cite{Yu:2017oky}. Unlike the SCS processes, in which the precise measurement of the $CP$ asymmetry cannot discriminate
new physics due to the ambiguities in estimating the penguin amplitudes, the measurable direct $CP$ asymmetry in $\mathcal{B}_c\to \mathcal{B}K(t)(\to \pi^+\pi^-)$ indicates new physics effect because of its tiny value in the SM.

The time-integrated $CP$ asymmetry is covered by the bare asymmetry with a time-dependent function, $F(t)$, introduced to take into account relevant experimental effects,
\begin{equation}\label{z21}
 A_{CP}(0,\infty)= \frac{\int^\infty_0dt\,F(t)\,\big[A_{CP}^{\overline K^0}(t)+A_{CP}^{\text{dir}}(t)+A_{CP}^{\text{int}}(t)\big]}
 {\int^\infty_0dt\,F(t)\,D(t)}.
\end{equation}
In this work, we adopt the approximation in \cite{Grossman:2011zk}
\begin{equation}
  F(t)=\begin{cases}&1 ~~~~~ t_1<t<t_2,\\
  &0~~~~~t>t_2~~\text{or}~~t<t_1.
\end{cases}
\end{equation}
Eq. \eqref{z21} is reduced as
\begin{align}\label{int}
A_{CP}(t_1,t_2)&\simeq \frac{2r_f\sin\delta_f\sin\phi}{1-2r_f\cos\delta_f\cos\phi} + \frac{2\mathcal{R}e[\epsilon]-4\mathcal{I}m[\epsilon]r_f\cos\phi\sin\delta_f}{1-2r_f\cos\delta_f\cos\phi}\nonumber\\&~~~~~~~ \times \Bigg[1- \frac{\big[c(t_1)-c(t_2)\big]+\frac{\mathcal{I}m[\epsilon]+2\mathcal{R}e[\epsilon]r_f\cos\phi\sin\delta_f}{\mathcal{R}e[\epsilon]-2\mathcal{I}m[\epsilon]r_f\cos\phi\sin\delta_f}
\big[s(t_1)-s(t_2)\big]}{\tau_S\Gamma \,(1+x^2)(e^{-t_1/\tau_S}-e^{-t_2/\tau_S})}\Bigg],
 \end{align}
in which $x\equiv\Delta m/\Gamma$, $c(t)=e^{- \Gamma t}[\cos(\Delta m t)-x\, \sin(\Delta m t)]$, and $s(t)=e^{- \Gamma t}[x \cos(\Delta m t)+ \sin(\Delta m t)]$.
The first term, being independent of $t_{1,2}$, corresponds to the direct $CP$ asymmetry in charm decays. In the rest part of Eq. \eqref{int}, the terms proportional to $r_f$ represent the new effect $A^{\rm int}_{CP}(t_1,t_2)$, and those without $r_f$ are the $CP$ violation in the neutral kaon mixing.

In the limitation of $t_1\ll \tau_S\ll t_2 \ll \tau_L$, $e^{-\Gamma t_1}=e^{-\Gamma_S t_1}=1$ and $e^{-\Gamma t_2}=e^{-\Gamma_S t_2}=0$. The time-integrated $CP$ violation then reads as
\begin{align}
A_{CP}(t_1\ll \tau_S\ll t_2 \ll \tau_L)&\simeq \frac{2r_f\sin\delta_f\sin\phi}{1-2r_f\cos\delta_f\cos\phi}+ \frac{2(\mathcal{R}e[\epsilon]-2\mathcal{I}m[\epsilon]r_f\cos\phi\sin\delta_f)}{1-2r_f\cos\delta_f\cos\phi}\nonumber\\&~~~~~\times~~\Bigg[1-
\frac{2}{1+x^2}-\frac{\mathcal{I}m[\epsilon]+2\mathcal{R}e[\epsilon]r_f\cos\phi\sin\delta_f}
{\mathcal{R}e[\epsilon]-2\mathcal{I}m[\epsilon]r_f\cos\phi\sin\delta_f}\frac{2x}{1+x^2}\Bigg].
 \end{align}
Under the approximation of $\mathcal{I}m[\epsilon] / \mathcal{R}e[\epsilon] \simeq -x/y$ and $y\approx-1$ \cite{Grossman:2011zk}, we get
\begin{align}\label{x1}
 A_{CP}(t_1\ll \tau_S\ll t_2 \ll \tau_L)& \simeq
  \frac{-2\mathcal{R}e[\epsilon]+2r_f\sin\delta_f\sin\phi-4r_f\mathcal{I}m[\epsilon]\cos\phi\sin\delta_f}
 {1-2r_f\cos\phi\cos\delta_f}\nonumber\\& = \big[A_{CP}^{\overline K^0}+A_{CP}^{\text{dir}}+A_{CP}^{\text{int}}\big]/D.
\end{align}
Considering the sizes of $\epsilon$, $r_f$ and $\phi$, the interference between charm decays and neutral kaon mixing $A_{CP}^{\text{int}}$ is expected to be of the order of $\mathcal{O}(10^{-4\sim-3})$. Such an order is much larger than the direct $CP$ violation in charm decays.

In Eq. \eqref{x1}, the $2r_f\cos\phi\cos\delta_f$ term in the denominator cannot be neglected, since the new $CP$-violation effect, $4r_f\mathcal{I}m[\epsilon]\cos\phi\sin\delta_f$, is a sub-leading contribution which is at the same order as the term in the denominator by expansion to be $4r_f\mathcal{R}e[\epsilon]\cos\phi\cos\delta_f$. The $2r_f\cos\phi\cos\delta_f$ term can be determined by the measurement of the $K^0_S-K^0_L$ asymmetry since the weak phase $\phi$ is tiny in the SM,
\begin{equation}\label{eq:denoR}
 R(\mathcal{B}_c\to \mathcal{B}K^0_{S,L})\simeq -2r_f\cos\delta_f \simeq -2r_f\cos\phi\cos\delta_f.
\end{equation}
In the case of $D^+\to \pi^+K^0_S$ decay, the term in the denominator is one order of magnitude smaller than the new $CP$-violation effect and can be neglected \cite{Yu:2017oky}, since $R(D^+\to \pi^+K^0_S)$ is measured to be small \cite{He:2007aj}
\begin{equation}
  R(D^+\to \pi^+K^0_{S,L})=0.022\pm0.016\pm0.018.
\end{equation}
But in the case of $\Lambda^+_c\to pK^0_S$, the contribution from the denominator can only be determined by the measurement of the $K_S^0-K_L^0$ asymmetry.

The time-dependent and time-integrated $CP$ violation in $\Lambda^+_c\to pK^0_S$ can be measured by Belle II and LHCb.
In order to extract the new $CP$-violation effect $A_{CP}^{\rm int}$, we
propose an observable,
\begin{align}
  \Delta A_{CP}(K^0_S,p,\pi^+)  &\equiv  A_{CP}(\Lambda^+_c\to pK^0_S)-A_{CP}(D^+\to \pi^+K^0_S) \nonumber\\
   & ~~\simeq A_{CP}^{\rm int}(\Lambda^+_c\to pK^0_S)-A_{CP}^{\rm int}(D^+\to \pi^+K^0_S)\nonumber\\
   & ~~~~~~ -A_{CP}^{\overline K^0}\cdot\big[ D(\Lambda^+_c\to pK^0_S)-D(D^+\to \pi^+K^0_S)\big],
\end{align}
where $D$ in the last line is the denominator of Eq.(\ref{x1}).
The $CP$ violation in neutral kaon mixing is mode-independent and thus cancelled between the ones in $\Lambda^+_c\to pK^0_S$ and $D^+\to \pi^+K^0_S$.  As discussed above, the direct $CP$ violation is negligible. Therefore, the new $CP$-violation effect can be revealed in $ \Delta A_{CP}(K^0_S,p,\pi^+)$. It can be measured at LHCb by combination of the raw asymmetries $A_{\rm raw}$ with those in CF channels, canceling the production and detection asymmetries, as
\begin{align}
  \Delta A_{CP}(K^0_S,p,\pi^+)  &=  [A_{\rm raw}(\Lambda^+_c\to pK^0_S)-A_{\rm raw}(\Lambda^+_c\to pK^-\pi^+)]\nonumber\\  & ~~~~~-[ A_{\rm raw}(D^+\to \pi^+K^0_S)-A_{\rm raw}(D^+\to K^-\pi^+\pi^+)].
\end{align}

\section{Numerical analysis}\label{re}

In the this section, we numerically estimate the $K^0_S-K^0_L$ asymmetries and $CP$ asymmetries in charmed baryon decays into neutral kaons based on the flavor $SU(3)$ symmetry.
The CF and DCS processes are model-independently analyzed with the parameters extracted from experimental data.
In this work, we focus on the decays of  the charmed-baryon anti-triplet ($\Xi_{c}^0$, $-\Xi_{c}^+$, $\Lambda_c^+$) into light baryon octet and pseudoscalar octet.

Note that sizable flavor $SU(3)$ breaking effect is always expected in charmed hadron decays. It is found to be very large in the SCS processes of $D$ decays, like the difference between $D^0\to K^+K^-$ and $\pi^+\pi^-$. However, the $SU(3)$ symmetry seems to work well in the CF processes. For example, in the analysis of $D$ meson decays using the topological diagrammatic approach under the $SU(3)$ symmetry in Ref. \cite{Cheng:2010ry}, the fitted results show that $\chi^2/d.o.f=  0.65$ for a global fit with only CF processes, while $\chi^2/d.o.f = 87$ in the case with only SCS processes.
    Since we focus on the charmed baryon decays into neutral kaons in the CF and DCS processes in this work, the SCS processes are not involved in the global fit. Thus large $SU(3)$ breaking effects can be avoided.
Besides, unlike the $D^0$ decays into neutral kaons in which the strong phase difference $\delta_f=0$ in the flavor $SU(3)$ limit \cite{Falk:1999ts,Wolfenstein:1995kv,Chau:1986du,Bhattacharya:2008ss,Wang:2017ksn,Jiang:2017zwr} and thus $A_{CP}^{\rm int}\propto\sin\delta_{f}$ is unobservable, $\delta_{f}$ and $A_{CP}^{\rm int}$ are non-vanishing in the charged $D$-meson decays \cite{Yu:2017oky,Li:2012cfa,Li:2013xsa}. In charmed baryon decays, it will be found in the following that the CF and DCS amplitudes are different in the $SU(3)$ decomposition, inducing non-vanishing $\delta_{f}$ and $A_{CP}^{\rm int}$.

The nonleptonic charmed decays are induced by the operators $(\bar s c)(\bar ud)$ for the CF modes,  and $(\bar d c)(\bar us)$ for the DCS modes. These operators can be decomposed into irreducible representations of flavor $SU(3)$ symmetry group. For example,
$( \bar s c )(\bar ud) = {\cal O}_6 + {\cal O}_{\overline {15}}$,
with
${\cal O}_6 = \frac{1}{2} [(\bar sc )(\bar ud)-(\bar u c )(\bar s d)]$ and
${\cal O}_{\overline {15}} = \frac{1}{2} [(\bar sc )(\bar ud)+(\bar u c )(\bar s d)]$.
The perturbative QCD corrections enhance the coefficient of ${\cal O}_6$ over ${\cal O}_{\overline {15}}$ by a factor of
 $\left[\alpha_s(m_b)/\alpha_s(m_W)\right]^{18/23}  \left[\alpha_s(m_c)/\alpha_s(m_b)\right]^{18/25}  \sim 2.5$ \cite{Gaillard:1974nj,Altarelli:1974exa}.
In an approximation of neglecting the contributions from $\mathcal{O}_{\overline{15}}$,
the effective Hamiltonian is expressed as \cite{Lu:2016ogy,Savage:1989qr,Geng:2017esc}
\begin{equation}\label{x2}
  \begin{split}
 \mathcal{H}_{\rm eff} =& e \mathcal{H}^{ab}(6) T_{ac} \overline {\mathcal{B}}^c_d M^d_b + f \mathcal{H}^{ab}(6) T_{ac} M^c_d \overline {\mathcal{B}}^d_b+ g \mathcal{H}^{ab}(6) \overline {\mathcal{B}}^c_a M^d_b T_{cd},
  \end{split}
\end{equation}
with the charmed-baryon anti-triplet
\begin{align}
T_{c}= (\Xi_{c}^0, -\Xi_{c}^+, \Lambda_c^+),\quad\quad T_{ab}=\epsilon_{abc}T^c,
\end{align}
the light baryon octet,
\begin{eqnarray}
\mathcal{B}^a_b=  \left( \begin{array}{ccc}
    \frac{1}{\sqrt 6} \Lambda + \frac{1}{\sqrt 2} \Sigma^0  & \Sigma^+  & p \\
    \Sigma^- &   \frac{1}{\sqrt 6} \Lambda - \frac{1}{\sqrt 2} \Sigma^0  & n \\
    \Xi^- & \Xi^0 & -\sqrt{2/3}\Lambda  \\
  \end{array}\right),
\end{eqnarray}
 and the pseudoscalar octet
 \begin{eqnarray}
 M^a_b=  \left( \begin{array}{ccc}
   \frac{1}{\sqrt 2} \pi^0+  \frac{1}{\sqrt 6} \eta_8    & \pi^+  & K^+ \\
    \pi^- &   - \frac{1}{\sqrt 2} \pi^0+ \frac{1}{\sqrt 6} \eta_8   & K^0 \\
    K^- & \bar K^0 & -\sqrt{2/3}\eta_8 \\
  \end{array}\right).
\end{eqnarray}
The non-zero elements $\mathcal{H}^{22}(6)=2$ for Cabibbo-allowed modes, and $\mathcal{H}^{33}(6)= 2 \tan^2\theta_C$ for doubly Cabibbo-suppressed modes, where $\theta_C$ is the Cabibbo angle and $\tan^2\theta_C\simeq |V_{cd}^*V_{us}/V_{cs}^*V_{ud}|$. The coefficients $e,\,f,\,g$ are free complex parameters to be extracted from data of branching fractions.

The  partial decay width of charmed baryon decays is
\begin{equation}\label{x6}
\Gamma(\mathcal{B}_c\rightarrow \mathcal{B}M) = \frac{|p_c|m_{\mathcal{B}}}{2\pi m_{\mathcal{B}_c}}|\mathcal{A}|^2,
\end{equation}
where $p_c$ is the center-of-mass momentum of the final state particles, $m_{\mathcal{B}_c}$ and $m_\mathcal{B}$ are the masses of charmed baryon and light baryons. The decay amplitudes are expressed by the $e$, $f$, $g$ parameters, with the representations given in Table \ref{tab:Br}. Since only the relative phases between $e$, $f$ and $g$ make sense, we take $e$ as real.  At the current stage, the available relevant data include five branching fractions of CF  decays of $\Lambda_c^+$ (shown in Table \ref{tab:Br}) and the ratio of branching fractions between $\Xi_{c}^0\to \Lambda K^0_S$ and $\Xi_{c}^0\to \Xi^-\pi^+$. So then five free parameters are fixed by six data via a global fit.
Notice that $f$ and $g$ always occur together as $(f+g)$ in the CF amplitudes, we use $h = f + g$ in the fit to avoid large correlation,
and $g$ can be obtained through $g=h-f$.  Then
we find
\begin{equation}\label{pa1}
  e = 0.67\pm 0.03,  \qquad|f| = 0.26\pm 0.03, \qquad h= f+g=\left( 0.43\pm 0.06\right)\,e^{i\left(0.97\pm 0.06\right)},
\end{equation}
with $\chi^2/d.o.f=0.17$, and the phase of $f$ ranges from $-\pi$ to $\pi$.
One can find the magnitudes of $|e|$ and $|f|$ are mostly fixed by $\Lambda_c^+\to pK_S^0$ and $\Xi^0K^+$, respectively, where the DCS contribution in the $pK_S^0$ mode is highly suppressed. The parameter $h$ is also well determined by global fitting. The phase of $f$ cannot be extracted from the available data, and thus is free in the full range between $-\pi$ and $\pi$.
The numerical results of branching fractions are given in the last column in Table \ref{tab:Br}, compared to the experimental data \cite{Patrignani:2016xqp}.
The uncertainties in our predictions include the errors from
the global fit and the ones from the lifetimes of $\Lambda^+_c$, $\Xi_c^+$ and $\Xi^0_c$.
For those modes in which $g$ occurs without $f$, we only give a range due to the ambiguity of the phase of $f$, including the uncertainties from the other fitted parameters and from the lifetimes of charmed baryons.
The exceptions in the DCS amplitudes do not affect the fitting very much. Besides, the branching fractions of $\Lambda_c^+\to \Sigma^0\pi^+$ and $\Sigma^+\pi^0$ are identical to each other due to the isospin symmetry.
From Table \ref{tab:Br}, it is clear that our results are well consistent with the data.  It is plausible that the CF and DCS transitions of charm baryons are well expressed by flavor $SU(3)$ symmetry.
     Besides, in Ref.~[27], the authors studied charmed baryon decays under the $SU(3)$ symmetry analysis similarly to our work, but including the SCS processes. Their $\chi^2$ is much larger than ours, and their result of $\mathcal{B}(\Lambda^+_c\to p\pi^0)$ significantly exceeds the experimental upper bound.
The predictions on the branching fractions of $\Lambda^+_c\to pK^0_L$ and those of $\Xi_c^{+,0}$ decays can be tested by experiments in the future.

\begin{table*}[t!]
\caption{Branching fractions and amplitude representations for the Cabibbo-favored charmed baryon decays. For the modes with neutral kaons $K_{S,L}^{0}$ in the final states, the associated doubly Cabibbo-suppressed amplitudes are taken into account with the factor of $\tan^{2}\theta_{C}$. The ratio of $BR(\Xi_{c}^0\to \Lambda K^0_S)/BR(\Xi_{c}^0\to \Xi^-\pi^+)=0.210\pm0.028$ is also included in the global fit. Our results are given in the last column, compared to the experimental data \cite{Patrignani:2016xqp}. }\label{tab:Br}
\centering
\small
\begin{tabular}{cccccc}
\hline\hline
 Modes &   ~Representation~    &
$BR_{\rm exp}(\%)$~& ~$BR_{\rm SU(3)}(\%)$ \\\hline
$\Lambda^+_c\to \Lambda \pi^+$ &  $\frac{1}{\sqrt{6}}(-2e-2f-2g)$   & 1.30$\pm $0.07
& 1.30$\pm 0.17$ \\
$\Lambda^+_c\to \Sigma^0\pi^+$ &$ \frac{1}{\sqrt{2}}(-2e+2f+2g)$  &1.29$\pm $0.07
& 1.27$\pm 0.17$  \\
$\Lambda^+_c\to \Sigma^+\pi^0$ &$ \frac{1}{\sqrt{2}}(2e-2f-2g)$  &1.24$\pm $0.10
& 1.27$\pm 0.17$  \\
$\Lambda^+_c\to pK^0_S$ &$ \frac{1}{\sqrt{2}}\tan^2\theta_C(2g)-\frac{1}{\sqrt{2}}(-2e)$ &1.58$\pm $0.08
& $1.36\sim 1.80$ \\
$\Lambda^+_c\to pK^0_L$ &$ \frac{1}{\sqrt{2}}\tan^2\theta_C(2g)+\frac{1}{\sqrt{2}}(-2e)$  &
 & $1.24\sim 1.67$  \\
$\Lambda^+_c\to \Xi^0K^+$ &$ -2f$  &0.50$\pm$0.12
 & 0.50$\pm 0.12$ \\
 \hline
$\Xi_{c}^0\to \Xi^-\pi^+$ &$ 2e$  &
& 2.24$\pm 0.34$ \\
$\Xi_{c}^0\to \Xi^0\pi^0$ &$ \frac{1}{\sqrt{2}}(-2e+2g)$  &
& $0.07\sim 1.81$ \\
$\Xi_{c}^0\to \Lambda K^0_S$ &$ \frac{1}{\sqrt{12}}\tan^2\theta_C(-2e+4f+4g)-\frac{1}{\sqrt{12}}(-4e+2f+2g)$  &
& 0.47$\pm 0.08$ \\
$\Xi_{c}^0\to \Lambda K^0_L$ &$ \frac{1}{\sqrt{12}}\tan^2\theta_C(-2e+4f+4g)+\frac{1}{\sqrt{12}}(-4e+2f+2g)$  &
& 0.50$\pm 0.09$  \\
$\Xi_{c}^0\to \Sigma^+K^-$ &$ 2f$  &
& 0.31$\pm 0.09$\\
$\Xi_{c}^0\to \Sigma^0 K^0_S$ &$ \frac{1}{2}\tan^2\theta_C(2e)-\frac{1}{2}(-2f-2g)$  &
& 0.23$\pm 0.07$   \\
$\Xi_{c}^0\to \Sigma^0 K^0_L$ &$\frac{1}{2}\tan^2\theta_C(2e)+\frac{1}{2}(-2f-2g)$  &
& 0.20$\pm 0.06$  \\
  \hline
$\Xi_{c}^+\to \Xi^0\pi^+$ &$ -2g$  &
& $0.01\sim 10.22$    \\
$\Xi_{c}^+\to \Sigma^+ K^0_S$ &$ \frac{1}{\sqrt{2}}\tan^2\theta_C(-2e)-\frac{1}{\sqrt{2}}(2g)$  &
& $0.06\sim 4.84$ \\
$\Xi_{c}^+\to \Sigma^+ K^0_L$ &$\frac{1}{\sqrt{2}}\tan^2\theta_C(-2e)+\frac{1}{\sqrt{2}}(2g)$  &
& $0.00\sim 4.30$
\\\hline\hline
\end{tabular}
\end{table*}

From Table \ref{tab:Br}, one can find the branching fractions of $\mathcal{B}_c\rightarrow \mathcal{B}K_{S}^0$  are obviously
different from the $\mathcal{B}_c\rightarrow \mathcal{B}K_{L}^0$ ones due to the interference between the CF and DCS amplitudes, which results in measurable $K^0_S-K^0_L$ asymmetries.
The results of $K^0_S-K^0_L$ asymmetries, $R(\mathcal{B}_c\to \mathcal{B}K^0_{S,L})$, are given in in Table~\ref{tab:para}.
It is found that the $K^0_S-K^0_L$ asymmetry in $\Lambda^+_c\to pK^0_{S,L}$ decay, the promising observable to search for two-body DCS amplitudes of charmed baryons, can reach the order of $10^{-2}$ or even $0.1$, within the experimental capability.

\begin{table*}[t!]
\caption{$K^0_S-K^0_L$ asymmetries in $\mathcal{B}_c\to \mathcal{B}K^0_{S,L}$ decays.}\label{tab:para}
\centering
\small
\begin{tabular}{cccc}
\hline\hline
 ~$R(\Lambda_c^+\to pK_{S,L}^{0})$~& ~$R(\Xi_{c}^0\to \Lambda K_{S,L}^{0})$~ & ~$R(\Xi_{c}^0\to \Sigma^0K_{S,L}^{0})$~ & ~$R(\Xi_{c}^+\to \Sigma^+K_{S,L}^{0})$~ \\\hline
 $-0.010\sim0.087$&$-0.037\pm 0.004$ & $0.091\pm 0.016$ & $-0.113\sim 0.390$
 \\\hline\hline
\end{tabular}
\end{table*}

The numerical results of the time-integrated $CP$ asymmetries $A_{CP}(t_1\ll\tau_S\ll t_2 \ll \tau_L)$ (denoted by $A_{CP}$ for simplification) are presented in Table~\ref{tab:acp}. There are two solutions of the $CP$ asymmetries for each parameter set.
We label the results obtained from Eq.~\eqref{pa1} by $S1$, and $S2$ is obtained by flipping the signs of all phase parameters in $S1$ since
the solution with opposite strong phases contributes equivalently to branching fractions
 that are proportional to the cosine of the strong phases.
The measurements in the future, to establish the
above $CP$ asymmetries, could discriminate these solutions.
In the absence of the DCS contributions, i.e., $r_f=0$, $A_{CP}(\mathcal{B}_c\to \mathcal{B}K^0_S) = A_{CP}^{\overline K^{0}}\simeq -2\mathcal{R}e[\epsilon]\approx -3.23\times 10^{-3}$. The new $CP$-violation effect $A_{CP}^{\rm int}$ can be revealed by the subtraction of  $A_{CP}^{\overline K^{0}}$ and and the $-2r_f\cos\phi\cos\delta_f$ term in denominator of Eq. \eqref{x1} from the total $CP$ asymmetries.
The signs of $A_{CP}^{\rm int}$ are opposite between the solutions of $S1$ and $S2$ due to its proportion to $\sin\delta_f$, while the ones of $-2r_f\cos\phi\cos\delta_f$ are the  same in solutions of $S1$ and $S2$.
The denominator $D$ in Eq. \eqref{x1} could be obtained from the corresponding $K^0_S-K^0_L$ asymmetries, seen in Eq. (\ref{eq:denoR}).
The new $CP$ asymmetry effect, $A_{CP}^{\text{int}}$, is of the order of $10^{-4}$, while the direct $CP$ asymmetry $A_{CP}^{\rm dir}$ is $\mathcal{O}(10^{-5})$.
For example, the range of $ A_{CP}^{\text{int}}(\Lambda_c^+\to pK_{S}^{0})$ is $(0.16\sim 3.37)\times 10^{-4}$ and the one for $ A_{CP}^{\text{dir}}(\Lambda_c^+\to pK_{S}^{0})$ is $(0.3\sim 6.8)\times 10^{-5}$ in solution $S1$.
The direct $CP$ violation are sensitive to new physics which may contribute to the DCS amplitudes with a large weak phase \cite{Yu:2017oky}. Thus if the new $CP$ violating effect was determined in experiment, the direct $CP$ violation in charmed baryon decays into neutral kaons can be obtained and used to search for new physics.

\begin{table*}[t!]
\caption{ Time-integrated $CP$ asymmetries $A_{CP}(t_1\ll\tau_S\ll t_2 \ll \tau_L)$ in $\mathcal{B}_c\to \mathcal{B}K^0_{S}$ decays in the units of $10^{-3}$, where the sets of $S2$ are obtained by flipping the signs of all phase parameters of $S1$.}\label{tab:acp}
\centering
\small
\begin{tabular}{ccccc}
\hline\hline
 & $A_{CP}(\Lambda_c^+\to pK_{S}^{0})$ & $A_{CP}(\Xi_{c}^0\to \Lambda K_{S}^{0})$ & $A_{CP}(\Xi_{c}^0\to \Sigma^0K_{S}^{0})$ & $A_{CP}(\Xi_{c}^+\to \Sigma^+K_{S}^{0})$ \\
\hline
$S1$ &$-3.15\sim-2.67$ &$-3.13\pm 0.05$ &$-3.42\pm 0.05$ &$-4.57\sim -2.60$ \\
  $S2$ & $-3.55\sim-3.09$ & $-3.58\pm 0.04$  & $-2.50\pm 0.10$ & $-2.91\sim-1.39$\\
  \hline\hline
\end{tabular}
\end{table*}

\section{Summary}\label{co}

Charmed baryon decays are becoming more intriguing due to the new measurements by Belle and BESIII in recent years. In this work, we investigate the $K^0_S-K^0_L$ asymmetries and $CP$ violation in charmed baryon decays into neutral kaons. Induced by the interference between the Cabibbo-favored and the doubly Cabibbo-suppressed amplitudes, the $K_{S}^{0}-K_{L}^{0}$ asymmetries can be used to study the DCS amplitudes. As no evidence of two-body DCS process in charmed baryon decays has been found so far, we propose to measure the $K^0_S-K^0_L$ asymmetry in the $\Lambda^+_c\to pK^0_{S,L}$ decay mode as a promising method to search for the two-body DCS transition. Besides, a new $CP$-violation effect is found in charmed baryon decays into neutral kaons, induced by the interference between the CF and DCS amplitudes with the $K^0-\overline K^0$ mixing. Once it is determined in experiments, the direct $CP$ asymmetries can be used to search for new physics beyond the Standard Model.
A numerical analysis based on $SU(3)$ symmetry is preformed to estimate the values of $K^0_S-K^0_L$ asymmetries and $CP$ violations.

\acknowledgments

We are grateful to Lei Li and Xiao-Rui Lyu for helpful discussions.
This work was supported in part by the National Natural Science Foundation of China under
Grants No. 11347027, 11375076, 11505083 and U1732101, and the Fundamental Research Funds for the
Central Universities under Grant No. lzujbky-2015-241 and lzujbky-2017-97.

\end{document}